\documentclass[conference,letterpaper]{IEEEtran}

\usepackage[left = 1.4cm, right =1.4cm, top = 1.4cm, bottom = 1.4cm]{geometry}
\usepackage[utf8]{inputenc} 
\usepackage[T1]{fontenc}
\usepackage{url}
\usepackage{ifthen}
\usepackage{cite}
\usepackage[cmex10]{amsmath}
\usepackage{cite}
\usepackage{amsmath,amssymb,amsfonts}
\usepackage{algorithmic}
\usepackage{graphicx}
\usepackage{makecell}
\usepackage{textcomp}
\usepackage{xcolor}
\usepackage{caption}
\usepackage{mathtools}
\usepackage{comment}
\newtheorem{thm}{Theorem}

\newtheorem{prop}{Proposition}

\newtheorem{exmp}{Example}

\DeclarePairedDelimiter\ceil{\lceil}{\rceil}
\DeclarePairedDelimiter\floor{\lfloor}{\rfloor}

\begin{document}
\title{Multi-Access Coded Caching with Secure Delivery} 

\author{\IEEEauthorblockN{K. K. Krishnan Namboodiri and B. Sundar Rajan}
	\IEEEauthorblockA{\textit{Department of Electrical Communication Engineering} \\
		\textit{Indian Institute of Science}\\
		Bengaluru, India \\
		\{krishnank, bsrajan\}@iisc.ac.in}
}

\maketitle

\begin{abstract}
The multi-access variant of the coded caching problem in the presence of an external wiretapper is investigated . A multi-access coded caching scheme with $K$ users, $K$ caches and $N$ files, where each user has access to $L$ neighbouring caches in a cyclic wrap-around manner, is proposed, which is secure against the wiretappers. Each transmission in the conventional insecure scheme will be now encrypted by a random key. The proposed scheme uses a novel technique for the key placement in the caches. It is also shown that the proposed secure multi-access coded caching scheme is within a constant multiplicative factor from the information-theoretic optimal rate for $L\geq \frac{K}{2}$ and $N\geq 2K$. 
\end{abstract}
\section{Introduction}
The technique of coded caching introduced in \cite{MaN} helps in reducing the peak-hour network traffic. This is achieved by making a part of the content locally available at the user end during off-peak hours. The proposed scheme in \cite{MaN} consists of a central server, having a library of $N$ files, connected to $K$ users through an error-free broadcast link. Each user is equipped with a dedicated cache, which can store $M$ out of the $N$ files in the placement phase. Each user reveals the demand in the delivery phase, which is assumed to be a single file from the $N$ possible choices. Then, the server broadcasts coded symbols to all the users over the shared link. The objective is to jointly design the placement and the delivery phases such that the load of the shared link in the delivery phase is minimized. 

However, in practical scenarios such as in cellular networks, users can have access to multiple caches when their coverage areas overlap. Incorporating this possibility, coded caching problem has been extended to multi-access set-up recently \cite{SPE,ReK,ReK2,CLWZC,SaR,MaR,SaR2}, where each user can access $L$ neighbouring caches in a cyclic wrap-around fashion. In \cite{SPE}, the authors give an optimal multi-access coded caching scheme when $\frac{M}{N}=\frac{K-1}{KL}$. A connection between index coding and multi-access coded caching is established in \cite{ReK2}. The construction of the multi-access schemes from placement delivery arrays (PDAs) is shown in \cite{CLWZC} and \cite{SaR}. In \cite{SaR,MaR,SaR2}, the authors construct multi-access coded caching schemes with linear subpacketization.

One of the main challenges of the coded caching problem is associated with the security of the multicast transmission that arises due to the broadcast nature of the channel. In \cite{STC}, the authors study the security aspects of the coded caching problem assuming each user is equipped with a dedicated cache. The scheme in \cite{STC} characterizes the fundamental limits of secure coded caching problem in the presence of a wiretapper. In \cite{RPKP}, the secret sharing technique is used to ensure that no user will be able to obtain any information, from its cache content as well as the server transmission, about any file other than the one it has requested. The coded caching schemes preserving privacy for user demands are studied in \cite{WaG,Kam,AST,GRKDK,NaR,YaT}. 

In this paper, we study the coded caching problem with secure delivery in a multi-access coded caching scheme. We extend the model proposed in \cite{STC} for the dedicated cache networks to the multi-access coded caching scheme, where an external wiretapper can listen to the transmissions in the delivery phase. The main contributions of this paper can be summarized as follows,
\begin{itemize}
	\item A secure multi-access coded caching scheme that is robust against external wiretappers is proposed (Section \ref{Results}, Theorem \ref{Thm:Scheme} and Section \ref{Scheme}).
	\item A novel technique for the key placement in the multi-access setting is introduced.
	\item The proposed scheme is shown to be optimal within a constant multiplicative factor for $L\geq \frac{K}{2}$ and $N\geq 2K$ (Section \ref{Results}, Theorem \ref{OptimalityGap}). 
\end{itemize}

\subsection{Notations}
For a positive integer $n$, $[n]$ denotes the set $ \left\{1,2,\hdots,n\right\}$.
For any two integers, $i$ and $K$, 
$$
<i>_K =
\begin{cases}
	i\text{ }(mod\text{ }K) & \text{if $i$ $(mod$ $K) \neq0$. }\\
	K & \text{if $i$ $(mod$ $K) =0$.}
\end{cases}      
$$
$gcd(a,b)$ represents the greatest common divisor of two positive integers $a$ and $b$. The notation '$\oplus$' represents the element-wise Exclusive OR (XOR) operation between two binary vectors.
\section{preliminaries}
\subsection{Single Unicast Index Coding Problem with Symmetric and Consecutive side information (SUICP(SC))}

A single unicast index coding problem (SUICP) consists of the following \cite{YBJK}:
\begin{itemize}
	\item A sender with $K$ independent messages $\mathcal{M} = \{x_1,x_2,\hdots,x_K\}$, $x_i \in \{0,1\}$ for all $i\in [K]$.
	\item $K$ receivers, denoted by $\{R_1,R_2,\hdots,R_K\}$. Receiver $R_i$ wants the message $x_i$ and knows a set $\mathcal{K}_i\subseteq \mathcal{M}\backslash \{x_i\}$ a priori, called as the \textit{side information set} of $R_i$. The sender knows the side information set of all the receivers.  
	\item An encoding function for the sender, 
	$E_{ic} : \{0,1\}^K \longmapsto \{0,1\}^\ell,$
	where $\ell$ is the length of the index code.
	\item $K$ decoding functions, one for every user,
	$D_{ic}^{(k)} : \{0,1\}^\ell \times \{0,1\}^{|\mathcal{K}_k|} \longmapsto \{0,1\} \hspace{0.5cm} \forall k\in [K],$
	such that $D_{ic}^{(k)}(E_{ic}(\mathcal{M}),\mathcal{K}_k) = x_k$, for each receiver $ k$.
	
\end{itemize}
\noindent The broadcast rate $\ell^*$ of an index coding problem is the minimum number of index code symbols to be transmitted such that every receiver can decode its required message by using the broadcasted index code symbols and the available  side-information set.\\
\noindent A scalar linear index code of length $\ell$ is described by a matrix $A\in \mathbb{F}_2^{K\times \ell}$ and the broadcast codeword $\mathbf{c}$ is given as,
\begin{equation*}
	\mathbf{c}= \mathbf{x}A = \sum_{k=1}^K x_kA_k,
\end{equation*}
where $A_k$ is the $k^{th}$ row of matrix $A$.

In an SUICP(SC) with $K$ messages and $K$ receivers, each receiver will have $K-(U+D+1)$ consecutive messages in the side information set. That is, for receiver $k$, $
	\mathcal{K}_k = \{x_{<k+D+1>_K},x_{<k+D+2>_K},\hdots,x_{<k-U-1>_K}\}$,  
where $U$ and $D$ are two non-negative integers such that $U+D<K$. For a given $K,U$ and $D$, $r_K(U,D)$ represents an achievable index code length for the SUICP(SC) problem.
\subsection{Adjacent Independent Row (AIR) Matrices}     
For two positive integers $m,n$ and $m\geq n$, an AIR matrix $A_{m\times n}$ is a zero-one matrix with the property that any $n$ adjacent rows (with cyclic wrap-around) of $A$ are linearly independent over $\mathbb{F}_q$ (irrespective of the field size $q$). For example,   
\begin{equation*}
	\label{AIRmatrix}
	A_{5\times 3} = 
	\begin{bmatrix} 
		1 & 0 & 0 \\
		0 & 1 & 0 \\
		0 & 0 & 1  \\
		1 & 0 & 1  \\
		0 & 1 & 1  
	\end{bmatrix}.
\end{equation*}
It can be verified that any 3 adjacent rows of $A$ are linearly independent. A general construction of AIR matrices is given in \cite{VaR}. 

Consider an SUICP(SC) with $K$ receivers. Receiver $R_k$ wants the message $x_k$. $R_k$ does not have access to the next $D$ messages $\{x_{<k+1>_K},x_{<k+2>_K},\hdots,x_{<k+D>_K}\}$ and  the previous $U$ messages $\{x_{<k-1>_K},x_{<k-2>_K},\hdots,x_{<k-U>_K}\}$. Encoding $\mathbf{x} = [x_1,x_2,\hdots,x_K]$ with the AIR matrix $A_{K\times (U+D+1)}$ gives an index code with length $r_K(U,D)=U+D+1$ \cite{VaR}. 

\section{System Model}    
A central server having a library of $N$ independent files, $\mathbf{W}=\left\{W_1,W_2,\hdots,W_N\right\}$ each of size $F$ bits ($W_n \sim unif[\mathbb{F}_2^F]$), connected to $K$ users $\left\{U_1,U_2,\hdots,U_K\right\}$  through an error-free broadcast link. The system consists of $K$ caches $\left\{\mathcal{Z}_1,\mathcal{Z}_2,\hdots,\mathcal{Z}_K\right\}$ each of capacity $MF$ bits, where $0\leq M \leq N$. Each user has access to $L<K$ caches in a cyclic wrap-around manner. $\mathcal{L}_k$ denotes the caches accessible for user $U_k$, and $\mathcal{L}_k = \left\{\mathcal{Z}_k,\mathcal{Z}_{<k+1>_K},\hdots,\mathcal{Z}_{<k+L-1>_K}\right\}$. Note that, $|\mathcal{L}_k|=L, \text{ }\forall k\in [K]$. A system under the aforementioned setting is called a $(K,L,N)$ multi-access network. The coded caching schemes under this model have been discussed in \cite{SPE,ReK,ReK2,CLWZC,SaR,SaR2,MaR}.  

A $(K,L,N)$ multi-access coded caching scheme works in two phases. In the \textit{placement phase}, server fills the caches with the file contents, without knowing the user demands. $Z_k$ denotes the content stored in the cache $\mathcal{Z}_k$. In the \textit{delivery phase}, each user requests a single file from the server. Let $W_{d_k}$ be the file demanded by user $k$. Corresponding to the demand vector $\mathbf{d} = \{d_1,d_2,\hdots,d_K\}$, the server makes a transmission $X^{\mathbf{d}}$ of size $RF$ bits. Each user $U_k$, $k\in [K]$ should be able to decode the demanded file from the transmission $X^{\mathbf{d}}$ and the contents in the caches in $\mathcal{L}_k$. That is,
\begin{equation}
	\label{Decodability}
	\text{[Decodability]     } H(W_{d_k}|Z_{\mathcal{L}_k},d_k,X^{\mathbf{d}})=0,\text{  } \forall k\in [K].
\end{equation}

\begin{figure}[t]
	\begin{center}
		\captionsetup{justification = centering}
		\includegraphics[width = \columnwidth]{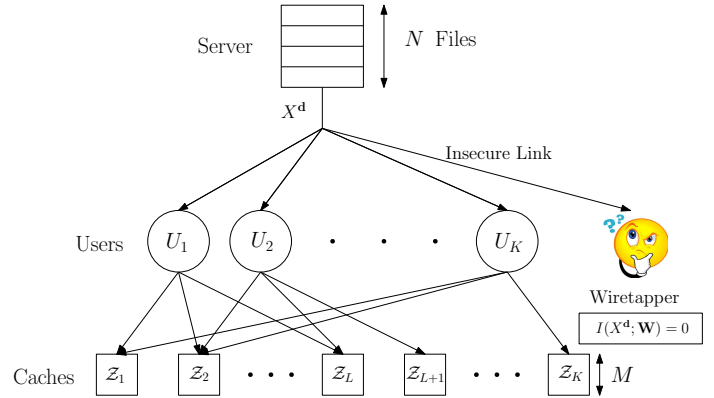}
		\caption{$(K,L,N)$ Multi-access Coded Caching Network in the presence of an external wiretapper. }
		\label{MACCNetwork}
	\end{center}
\end{figure}

In this work, we assume that the communication link from the server to the users is insecure, and we incorporate the possibility of an external wiretapper in the delivery phase, which can observe the broadcast transmission $X^{\mathbf{d}}$ (See Fig. \ref{MACCNetwork}). In addition to the decodability condition, we require that the communication $X^{\mathbf{d}}$ should not reveal any information about the files $\mathbf{W}$. That is,
\begin{equation}
	\label{Security}
	\text{[Security Condition]  }\hspace{0.5cm} I(X^{\mathbf{d}};\mathbf{W})=0.
\end{equation} 
For a $(K,L,N)$ multi-access coded caching scheme, a memory-rate pair $(M,R_s)$ is said to be securely achievable if there exists a scheme (for cache memory $M$ with rate $R_s$) satisfying the decodability condition \eqref{Decodability} and the security condition \eqref{Security} for every possible demand vectors. When $L=1$, the problem reduces to the setting in \cite{STC}. The multi-access coded caching schemes in \cite{SPE,ReK,ReK2,CLWZC,SaR,SaR2,MaR} do not satisfy the security condition.
\section{Main Results}
\label{Results}
In this section, we present a $(K,L,N)$ secure multi-access coded caching scheme by distributed key placement. Then we show that the proposed scheme is optimal within a constant gap when $L\geq \frac{K}{2}$. 
\begin{thm}
	\label{Thm:Scheme}
For a $(K,L,N\geq K)$ multi-access coded caching scheme, the following rates are securely achievable.
\begin{itemize}
	\item At $M=1$, $R_s =K$.
	\item If $i\in \{1,2,\hdots, \floor{\frac{N}{L}}\}$ and $gcd(i,K)=1$, then, at $M = \frac{iN}{K}+\frac{K}{L}(1-\frac{iL}{K})^2$, $R_s =K(1-\frac{iL}{K})^2$.
	\item If $i\in \{1,2,\hdots, \floor{\frac{N}{L}}\}$ and $g = gcd(i,K) \neq 1$, then, at $M = \frac{iN}{K}+\frac{2K}{g(L+1)}(1-\frac{iL}{K})^2$, $R_s =K(1-\frac{iL}{K})^2$.
	\item At $M = \frac{N}{L}$, $R_s=0$.
\end{itemize}
The lower convex envelope of the above memory-rate pairs are also securely achievable.
\end{thm}
The scheme is presented in Section \ref{Scheme}.

The total memory $M$ is the sum of data memory $M_D$ and key memory $M_K$. For the secure coded caching scheme, the cache memory has to be at least a file size ($M\geq 1$). At $M=1$, the entire memory is dedicated for storing the keys ($M_K=1$ and $M_D=0$). At $M = \frac{iN}{K}+\frac{K}{L}(1-\frac{iL}{K})^2 $ with $gcd(K,i)=1$, $M_D = \frac{iN}{K}$ and $M_K = \frac{K}{L}(1-\frac{iL}{K})^2$. Similarly, at $M =\frac{iN}{K}+\frac{2K}{g(L+1)}(1-\frac{iL}{K})^2$ with $gcd(i,K)=g\neq 1$, $M_D =\frac{iN}{K}$ and $M_K = \frac{2K}{g(L+1)}(1-\frac{iL}{K})^2$. At these memory points, we stick to uncoded placement of file contents. At $M =\frac{N}{L}$, the complete cache memory is filled with the file content, and $M_K=0$. At this memory point, we make use of the coded placement of file contents. If we restrict to uncoded placement, $R_s=0$ is achievable at $M=\ceil{\frac{N}{L}}$.

\begin{exmp}
	In this example, we illustrate the achievable scheme in Theorem \ref{Thm:Scheme} for a $(3,2,N)$ multi-access coded caching problem for $M = \frac{N}{3}+\frac{1}{6}$. The cache placement is shown in Fig. \ref{MACCNetworkK=3}. In the conventional coded caching setting, the server would have sent $W_{d_1,3}\oplus W_{d_2,1}\oplus W_{d_3,2}$ for a demand vector $\mathbf{d} = [d_1,d_2,d_3]$. The technique for ensuring secrecy is one-time pad scheme\cite{Sha}. That is, every multicast transmission in the conventional (insecure) coded caching scheme will now be encrypted by a random key. So in our example, the transmission will be encrypted by a key, $\mathbb{K}_1$ of size $\frac{F}{3}$ bits, $\mathbb{K}_1 \sim unif\{\mathbb{F}_2^{F/3}\}$. But, $\mathbb{K}_1$ has to be accessible for all the 3 users. For that, the key will be split into 2 non-overlapping 'sub-keys', $\mathbb{K}_1^{(1)}$ and $\mathbb{K}_1^{(2)}$. Thus the actual key is the concatenation of the sub-keys, $\mathbb{K}_1 = [\mathbb{K}_1^{(1)} \text{ }\mathbb{K}_1^{(2)}]$. The cache key placement is shown in Fig. \ref{MACCNetworkK=3}. So, from the placement phase all the 3 users will get the key $\mathbb{K}_1$. Thus from the transmission, $W_{d_1,3}\oplus W_{d_2,1}\oplus W_{d_3,2}\oplus \mathbb{K}_1$, the users can subtract out the key and get back the actual transmission. The wiretapper, which doesn't have access to the key would gain no information regarding any of the file contents. The secure delivery is achieved by using an additional cache memory of $\frac{F}{6}$ bits. Notice that the extra memory needed for storing the keys will not scale up with the number of files with the server. Therefore, as the number of files increases, the fraction of additional memory needed for storing keys will decrease.     
	\begin{figure}[t]
		\begin{center}
			\captionsetup{justification = centering}
			\includegraphics[width = 0.9\columnwidth]{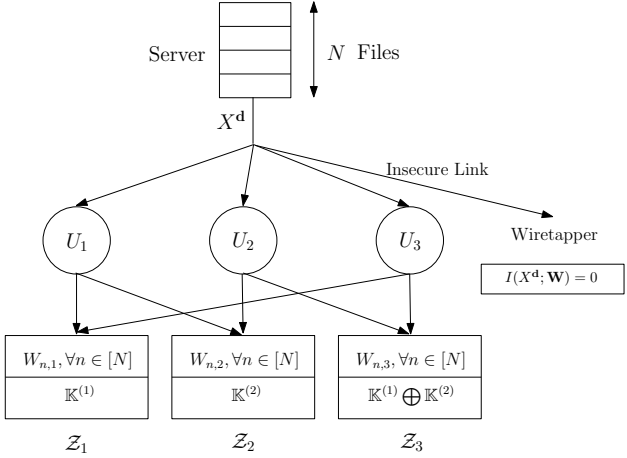}
			\caption{Cache Placement: $(K=3,L=2,N)$ Secure Multi-access Coded caching. }
			\label{MACCNetworkK=3}
		\end{center}
	\end{figure}
\end{exmp}

Now, we focus on the case with $L\geq \frac{K}{2}$ and restrict to the uncoded placement of file contents. In that case, the memory-rate pairs $(1,K)$, $(\frac{N}{K}+\frac{(K-L)^2}{KL}, K(1-\frac{L}{K})^2)$ and $(\frac{2N}{K},0)$ are securely achievable. Notice that, the key placement can be still coded. Let $R^*(M)$ denote the information-theoretic optimal memory-rate trade-off for the multi-access coded caching problem, restricted to the uncoded placement of file contents. In the following theorem, we show that the proposed scheme is order optimal when $N \geq 2K$ for $L\geq \frac{K}{2}$.
\begin{thm}
For a $(K,L,N)$ multi-access coded caching scheme with $L\geq \frac{K}{2}$ and $1\leq M\leq \frac{2N}{K}$,
\begin{equation*}
	\label{OptimalityGap}
		\frac{R_s(M)}{R^*(M)} \leq 
	\begin{cases}
		6 & \text{if $2K\leq N < 3K$. }\\
        4  & \text{if $N\geq 3K$.}
	\end{cases}   
\end{equation*}
\end{thm}
Proof of Theorem \ref{OptimalityGap} is given in Appendix \ref{appA}.
\begin{exmp}
	\label{Example2}
Now, we will illustrate one more example of the achievable scheme for $(K=6,L=2,N)$ with $i=2$. In the placement phase, the server divides each file into $\frac{K}{gcd(K,i)}=3$ subfiles, $W_n = \{W_{n,1},W_{n,2},W_{n,3}\}$. Then the server independently generates two random keys $\mathbb{K}_1,\mathbb{K}_2 \sim unif[\mathbb{F}_2^{F/3}]$. Split each key into $L+1 = 3$ sub-keys, $\mathbb{K}_i = [\mathbb{K}_i^{(1)}\text{ } \mathbb{K}_i^{(2)}\text{ } \mathbb{K}_i^{(3)}]$ for $i\in [2]$. The cache placement is summarized in Table \ref{CacheContent}. 
\begin{table}[h]
	\centering
	\begin{tabular}{|c|c|}
		\hline
		$\mathcal{Z}_1$ & $W_{n,1},\forall n\in [N]$, $\mathbb{K}_2^{(3)}$, $\mathbb{K}_1^{(1)}$\\
		\hline
		$\mathcal{Z}_2$ & $W_{n,2},\forall n\in [N]$, $\mathbb{K}_1^{(2)}$, $\mathbb{K}_1^{(3)}$\\
		\hline
		$\mathcal{Z}_3$ & $W_{n,3},\forall n\in [N]$, $\mathbb{K}_1^{(1)}$, $\mathbb{K}_1^{(2)}$\\
		\hline
		$\mathcal{Z}_4$ & $W_{n,1},\forall n\in [N]$, $\mathbb{K}_1^{(3)}$, $\mathbb{K}_2^{(1)}$\\
		\hline
		$\mathcal{Z}_5$ & $W_{n,2},\forall n\in [N]$, $\mathbb{K}_2^{(2)}$, $\mathbb{K}_2^{(3)}$\\
		\hline
		$\mathcal{Z}_6$ & $W_{n,3},\forall n\in [N]$, $\mathbb{K}_2^{(1)}$, $\mathbb{K}_2^{(2)}$\\
		\hline
	\end{tabular}
	\caption{Cache Placement}
	\label{CacheContent}
\end{table}
The cache memory $M$ is $\frac{N}{3}+\frac{2}{9}$. Let the demand vector be $\mathbf{d} = [d_1,d_2,d_3,d_4,d_5,d_6]$. Then, the transmission is,
\begin{equation*}
	\begin{split}
	X^{\mathbf{d}} = \big\{& W_{d_1,3}\oplus W_{d_2,1}\oplus W_{d_3,2}\oplus \mathbb{K}_1,  \\& W_{d_4,3}\oplus W_{d_5,1}\oplus W_{d_6,2}\oplus \mathbb{K}_2  \big\}.
    \end{split}
\end{equation*}
The transmission rate $R_s$ is $\frac{2}{3}$. Each user requires one subfile of their demanded file from the delivery phase. First 3 users can decode the intended subfiles from the transmission $ W_{d_1,3}\oplus W_{d_2,1}\oplus W_{d_3,2}\oplus \mathbb{K}_1$. Similarly, the users $U_4,U_5$ and $U_6$ can decode the subfile from the transmission $W_{d_4,3}\oplus W_{d_5,1}\oplus W_{d_6,2}\oplus \mathbb{K}_2$. Observe that the first 3 users need the key $\mathbb{K}_1$ and the remaining users need the key $\mathbb{K}_2$ for decrypting the transmission. From Table \ref{CacheContent}, it is clear that the users $U_1,U_2$ and $U_3$ has the key $\mathbb{K}_1$ from the placement phase. Similarly, the key $\mathbb{K}_2$ is available for the users $U_4,U_5$ and $U_6$.     
\end{exmp}
Before presenting the actual scheme, we show that for secure delivery, the cache memory $M\geq 1$.
\begin{prop}
	For a $(K,L,N)$ secure multi-access coded caching scheme with $N\geq K$, the cache memory $M\geq 1$. 
\end{prop}
\begin{IEEEproof}
For a demand vector $\mathbf{d} = [d_1,d_2,\hdots,d_K]$ with all $d_i$'s are distinct, we can write,
	\begin{equation}
		\label{NoUncertainty}
		H(W_{d_1},W_{d_2},\hdots,W_{d_K}|X^{\mathbf{d}},Z_1,Z_2,\hdots,Z_K) = 0.
	\end{equation}
\begin{equation}
	\label{NoInformation}
	I(W_{d_1},W_{d_2},\hdots,W_{d_K};X^{\mathbf{d}}) = 0
\end{equation}
\eqref{NoUncertainty} is a consequence of the decodability condition \eqref{Decodability}, and \eqref{NoInformation} is a consequence of the security condition \eqref{Security}. We have,
\begin{subequations}
	\begin{align}
		KF &= H(W_{d_1},W_{d_2},\hdots,W_{d_K})\\ 	
		   &=  I(W_{d_1},W_{d_2},\hdots,W_{d_K};   X^{\mathbf{d}},Z_1,\hdots,Z_K)\notag \\
		    &\qquad +H(W_{d_1},W_{d_2},\hdots,W_{d_K}| X^{\mathbf{d}},Z_1,\hdots,Z_K)\\
		   &= I(W_{d_1},W_{d_2},\hdots,W_{d_K}; X^{\mathbf{d}},Z_1,\hdots,Z_K)\\
		   &= I(W_{d_1},W_{d_2},\hdots,W_{d_K}; X^{\mathbf{d}})\notag \\
		   &\qquad +I(W_{d_1},W_{d_2},\hdots,W_{d_K};Z_1,\hdots,Z_K| X^{\mathbf{d}})\\
		   &= I(W_{d_1},W_{d_2},\hdots,W_{d_K};Z_1,\hdots,Z_K| X^{\mathbf{d}})\\
		   &\leq H(Z_1,\hdots,Z_K)\leq \sum_{k=1}^K H(Z_k) \leq KMF.
	\end{align}
\end{subequations} 
This implies $M\geq 1$.
\end{IEEEproof}

\section{Secure multi-access coded caching scheme}
In this section, we present the achievable scheme in Theorem \ref{Thm:Scheme}. 
\label{Scheme}
	\subsubsection{Case 1 ($M=1$)}
     At $M=1$, the entire memory is dedicated for key placement. That is, $M_K = 1$ and $M_D=0$. The server stores the random key $\mathbb{K}_k\sim unif[\mathbb{F}_2^F]$ in cache $\mathcal{Z}_k$, for all $k\in [K]$. For the demand vector $\mathbf{d} = \left[d_1,d_2,\hdots,d_K\right]$, the server will transmit $W_{d_k}\oplus \mathbb{K}_k$ for all $k\in [K]$. Therefore, $R_s = K$.
	
	\subsubsection{Case 2 ($gcd(K,i) =1$)}
    \textit{ a) Placement Phase}: The cache memory $M= \frac{Ni}{K}+\frac{K}{L}(1-\frac{Li}{K})^2$, where $M_D= \frac{Ni}{K}$ and $M_K = \frac{K}{L}(1-\frac{Li}{K})^2$. The cache file placement is as follows.
	 A file is divided into $K$ non-overlapping subfiles of same size, $F/K$ bits, and the $n^{th}$ file is thus $W_n=\left\{W_{n,1},W_{n,2},\hdots,W_{n,K}\right\}$. The server fills the cache $\mathcal{Z}_k$ with the subfiles $\{W_{n,<(k-1)i+1>_K},W_{n,<(k-1)i+2>_K},\hdots,W_{n,<ki>_K}\}$ for all $n\in[N]$. That is, content of $\mathcal{Z}_1$ is $\{W_{n,1},W_{n,2},\hdots, W_{n,i}\}$, $\mathcal{Z}_2$ stores $\{W_{n,<i+1>_K},W_{n,<i+2>_K},\hdots, W_{n,<2i>_K}\}$ and so on. By this placement strategy, it is ensured that each user has access to some $iL$ consecutive subfiles of all the files. That is, user $U_k$ has access to the subfiles stored in the set of caches $\mathcal{L}_k$. So $U_k$ gets $\{W_{n,<(k-1)i+1>_K},\hdots,W_{n,<(k+L-1)i>_K}\}$ for all $n\in [N]$ from the placement itself. Since $i \in \{1,2,\hdots,\floor{\frac{K}{L}}\}$, $K\geq iL$. Therefore, each user requires the remaining $K-iL$ subfiles of the demanded file in the delivery phase. 
	 
	 Now, for the cache key placement, the server independently generates $(K-iL)^2$ keys, $\mathbb{K}_\alpha\sim unif[\mathbb{F}_2^{F/K}]$, $\alpha \in [(K-iL)^2]$. Each key is divided into $L$ sub-keys each of size $\frac{F}{KL}$ bits. For all $\alpha \in [(K-iL)^2]$, the key $\mathbb{K}_\alpha = \{\mathbb{K}_{\alpha,1},\mathbb{K}_{\alpha,2},\hdots,\mathbb{K}_{\alpha,L}\}$. Consider the AIR matrix $A_{K\times L}$, where every adjacent $L$ rows of $A$ are linearly independent. Now encode $[\mathbb{K}_{\alpha,1},\mathbb{K}_{\alpha,2},\hdots,\mathbb{K}_{\alpha,L}]$ using $A$. That is,	$
		[\mathcal{K}_{\alpha,1},\mathcal{K}_{\alpha,2},\hdots,\mathcal{K}_{\alpha,K}]
		= [\mathbb{K}_{\alpha,1},\mathbb{K}_{\alpha,2},\hdots,\mathbb{K}_{\alpha,L}] A^T$. 
	\noindent The server fills the cache $k\in [K]$ with the coded sub-key $\mathcal{K}_{\alpha,k}$ of all the keys $\alpha \in [(K-iL)^2]$. The size of a coded sub-key is same as the size of a sub-key. Which means, $|\mathcal{K}_{\alpha,k}| = \frac{F}{KL}$ for all $k\in [K]$ and $\alpha\in[(K-iL)^2]$. Observe that, by the above key placement strategy, every user can get all the keys (by the linear independence property of the AIR matrix).

		\textit{b) Delivery Phase}: First we explain the delivery scheme without key encryption. Then to ensure secrecy, every transmission will be encrypted by a key. 
		 
		Let $W_{d_k}$ be the file demanded by user $U_k$ and the demand vector is thus $\mathbf{d} = [d_1, d_2,\hdots,d_K]$. As discussed previously, all users need $K-iL$ subfiles of their respective demanded file that are not available from the placement phase. This has been summarized in Table \ref{RequiredSubfiles}.	
		\begin{table*}[h]
			\centering
			\begin{tabular}{|c|c|c|c|c|c|c|}
				\hline
				&$U_1$ & $U_2$ & $\hdots$ & $U_k$ & $\hdots$ & $U_K$       \\
				\hline \hline
				row 1 & $W_{d_1,<iL+1>_K}$ & $W_{d_2,<i(L+1)+1>_K}$ &$\hdots$  & $W_{d_k,<i(L+k-1)+1>_K}$ &$\hdots$  & $W_{d_K,<i(L+K-1)+1>_K}$\\ \hline
				row 2 & $W_{d_1,<iL+2>_K}$ &$W_{d_2,<i(L+1)+2>_K}$ & $\hdots$ & $W_{d_k,<i(L+k-1)+2>_K}$ &$\hdots$  & $W_{d_K,<i(L+K-1)+2>_K}$\\ \hline
				\vdots &\vdots    & \vdots & \vdots & \vdots & \vdots & \vdots                             \\ \hline
				row j &$W_{d_1,<iL+j>_K}$ & $W_{d_2,<i(L+1)+j>_K}$ &$\hdots$  & $W_{d_k,<i(L+k-1)+j>_K}$ &$\hdots$  & $W_{d_K,<i(L+K-1)+j>_K}$\\ \hline
				\vdots &\vdots    & \vdots & \vdots & \vdots & \vdots & \vdots                             \\ \hline
				row $K-iL$ &$W_{d_1,K}$ & $W_{d_2,i}$& $\hdots$ & $W_{d_k,<i(k-1)>_K}$ &$\hdots$  & $W_{d_K,<i(K-1)>_K}$\\ \hline
				
			\end{tabular}
			\caption{Subfiles of the demanded file required by every user.}
			\label{RequiredSubfiles}
		\end{table*}
		Consider row $j$ of the Table \ref{RequiredSubfiles}. Since, $K$ and $i$ are relatively prime, $\{<i(k+L-1)+j>_K\}_{k=1}^K=\{1,2,\hdots,K\}$.
		\noindent Applying a permutation on row $j$,
		\begin{equation*}
			\begin{aligned}
				&\pi_j([W_{d_1,<iL+j>_K}, W_{d_2,<i(L+1)+j>_K},\hdots,\\&W_{d_K,<i(L+K-1)+j>_K}]) =[W_{d_{j_1},1},W_{d_{j_2},2},\hdots,W_{d_{j_K},K}]
			\end{aligned}
		\end{equation*}
		\noindent where $\pi_j(.)$ is the permutation operation on row $j$ and $j_k$ is $k$ such that $<i(k+L-1)+j>_K=k$. After the permutation operation $\pi_j(.)$, the $j_k^{th}$ user will be requiring the subfile $W_{j_k,k}$. But user $j_k$ has some $iL$ consecutive subfiles from $[W_{d_{j_1},1},W_{d_{j_2},2},\hdots,W_{d_{j_K},K}]$. That is, user $j_k$ has access to all the subfiles except those that are indexed with $\{<k-j+1>_K,<k-j+2>_K,\hdots,<k-1>_K\}\cup \{k\} \cup \{<k+1>_K,<k+2>_K,\hdots,<k+K-iL-j>_K\}$. The subfile required for user $j_k$ is $W_{d_{j_k},k}$. Thus the row $j$ corresponds to an index coding problem with symmetric and consecutive side information with $U=j-1$ and $D = K-iL-j$. By encoding the subfiles $[W_{d_{j_1},1},W_{d_{j_2},2},\hdots,W_{d_{j_K},K}]$ using a $K\times K-iL$ AIR matrix, we obtain an index coding solution of length $K-iL$. Same is repeated for all the $(K-iL)$ rows of Table \ref{RequiredSubfiles}. Therefore, there are $(K-iL)^2$ transmissions, and each transmission is encrypted by a key $\mathbb{K}_\alpha$. Thus the rate of transmission, $R_s=\frac{1}{K}(K-iL)^2 = K(1-\frac{iL}{K})^2$. The Decodability of the files is directly followed by the linear independence property of AIR matrices.
	\subsubsection{Case 3 ($gcd(K,i) \neq 1$)}
	
	Now, we will consider the case where $gcd(K,i)\neq 1$. Let $gcd(K,i) = g$. We define, $\tilde{K} := \frac{K}{g}$ and $\tilde{i} := \frac{i}{g}$. By definition, $gcd(\tilde{K},\tilde{i})=1$. 
	
	First we will see the multicast messages without key encryption. The server divides each file into $\tilde{K}$ non-overlapping subfiles of equal size, $F/\tilde{K}.$ The data memory $M_D = \frac{iN}{K} = \frac{\tilde{i}N}{\tilde{K}}$. The server fills cache $k$ with the subfiles $\{W_{n,<(k-1)\tilde{i}+1>_{\tilde{K}}},W_{n,<(k-1)\tilde{i}+2>_{\tilde{K}}},\hdots,W_{n,<k\tilde{i}>_{\tilde{K}}}\}$. That means, each cache contains $\tilde{i}$ consecutive subfiles of the total $\tilde{K}$ subfiles. By this placement strategy, the file contents stored in the caches $\{\mathcal{Z}_k,\mathcal{Z}_{\tilde{K}+k},\hdots,\mathcal{Z}_{(g-1)\tilde{K}+k}\}$, $k \in [\tilde{K}]$ are same. So, we are essentially splitting the entire users into $g$ mutually exclusive groups. The first group is the set of first $\tilde{K}$ users and second group is the next $\tilde{K}$ users and so on. Observe that each of these groups constitute an individual $(\tilde{K},L,N)$ multi-access coded caching scheme. So, there are $(\tilde{K}-\tilde{i}L)^2$ multicast messages corresponding to every group. Therefore, every user requires $(\tilde{K}-\tilde{i}L)^2$ keys. But there are $g$ groups. Hence the total number of keys that the server has to generate is $g(\tilde{K}-\tilde{i}L)^2$. The cache key placement is as follows.
	
	The server independently generates $g(\tilde{K}-\tilde{i}L)^2$ keys uniformly distributed over $\mathbb{F}_2^{F/\tilde{K}}$. Consider the keys $\mathbb{K}_1,\mathbb{K}_2,\hdots,\mathbb{K}_g$. Let the key $\mathbb{K}_j$ be used to encrypt a transmission corresponding to group $j\in [g]$. Divide each key into $L+1$ non-overlapping sub-keys, $\mathbb{K}_j^{(t)}, t \in [L+1]$ such that $\mathbb{K}_j = \{\mathbb{K}_{j,1},\mathbb{K}_{j,2},\hdots,\mathbb{K}_{j,L+1}\}$. Consider the vector $\big\{\mathbb{K}_1^{(1)},\mathbb{K}_1^{(2)},\hdots,\mathbb{K}_1^{(L+1)},\mathbb{K}_1^{(<L+2>_{L+1})},\hdots,$ $ \mathbb{K}_1^{(<2\tilde{K}>_{L+1})},\mathbb{K}_2^{(1)},\hdots,\mathbb{K}_2^{(<2\tilde{K}>_{L+1})},\mathbb{K}_3^{(1)},\hdots, \mathbb{K}_g^{(<2\tilde{K}>_{L+1})}\big\}$. The length of the vector is $2\tilde{K}g =2K$. Apply $L-1$ right cyclic shifts on the vector and store the first two elements of the resultant vector in $\mathcal{Z}_1$, the next two in $\mathcal{Z}_2$ and so on. For instance, in Example \ref{Example2}, $L=2$ and $g=2$. $U_1,U_2$ and $U_3$ are in first group and the remaining 3 users are in group 2. The keys $\mathbb{K}_1$ and $\mathbb{K}_2$ are divided into $L+1=3$ sub-keys. The vector $\{\mathbb{K}_1^{1},\mathbb{K}_1^{2},\mathbb{K}_1^{3},\mathbb{K}_1^{1},\mathbb{K}_1^{2},\mathbb{K}_1^{3},\mathbb{K}_2^{1},\mathbb{K}_2^{2},\mathbb{K}_2^{3},\mathbb{K}_2^{1},\mathbb{K}_2^{2},\mathbb{K}_2^{3}\}$ is subjected to a right shift operation by an amount $L-1=1$. The resulting vector is $\{\mathbb{K}_2^{3},\mathbb{K}_1^{1},\mathbb{K}_1^{2},\mathbb{K}_1^{3},\mathbb{K}_1^{1},\mathbb{K}_1^{2},$ $\mathbb{K}_1^{3},\mathbb{K}_2^{1},\mathbb{K}_2^{2},\mathbb{K}_2^{3},\mathbb{K}_2^{1},\mathbb{K}_2^{2}\}$. The sever stores $\mathbb{K}_2^{3}$ and $\mathbb{K}_1^{1}$ in $\mathcal{Z}_1$,  $\mathbb{K}_1^{2}$ and $\mathbb{K}_1^{3}$ in $\mathcal{Z}_2$ and so on. 
	
	By this key placement strategy, all the users in group $j$ obtain $\mathbb{K}_j^{(t)}$ for all $t\in [L+1]$. Because each user has access to $2L$ sub-keys. Since, we are applying $L-1$ right cyclic shifts, each user will get at least $L+1$ continuous sub-keys of the desired key. We showed the key placement of $g$ keys (one key per group). This will be repeated $(\tilde{K}-\tilde{i}L)^2$ times, so that each user will get $(\tilde{K}-\tilde{i}L)^2$ random keys (All the users in the same group will get the same $(\tilde{K}-\tilde{i}L)^2$ keys). Therefore $M_K = \frac{2\tilde{K}}{L+1}(1-\frac{iL}{K})^2$.
	
	In the delivery phase, once the demand of all the users are revealed, the previous delivery strategy can be used for each group separately since $\tilde{K}$ and $\tilde{i}$ are relatively prime (the transmissions intending to group $j\in [g]$ must be encrypted by the keys available for the users in group $j$). So, in this case the securely achieved rate is, $R_s =g\tilde{K}\left(1-\frac{\tilde{i}L}{\tilde{K}}\right)^2 = K\left(1-\frac{iL}{K}\right)^2$.
	\subsubsection{Case 4 ($M=\frac{N}{L}$)}
	
	At $\frac{M}{N} = \frac{1}{L}$, file $W_n$ is divided into $L$ non-overlapping subfiles of equal size, $F/L$. For all $n\in [N]$, the file $W_n = \{W_{n,1},W_{n,2},\hdots,W_{n,L}\}$. Consider the AIR matrix $A_{K\times L}$, every adjacent $L$ rows of $A$ are linearly independent. Now encode $[W_{n,1},W_{n,2},\hdots,W_{n,L}]$ using $A$. That is, $			[\mathcal{C}_{n,1},\mathcal{C}_{n,2},\hdots,\mathcal{C}_{n,K}]
		= [W_{n,1},W_{n,2},\hdots,W_{n,L}] A^T$. 
	The server fills the cache $k\in [K]$ with the coded subfile $\mathcal{C}_{n,k}$ of all the files $n\in [N]$. The size of a coded subfile is same as the size of a subfile. This means, $|C_{n,k}| = \frac{F}{L}$ for all $k\in [K]$ and $n\in[N].$ So, the memory constraint of all the caches are met by the given placement strategy. Each user can decode all the files from the accessible cache contents itself. User $U_k$ has access to $L$ adjacent caches in the cyclic wrap-around fashion. That means, $U_k$ possesses $L$ adjacent coded subfiles of every file. Since, every consecutive  $L$ columns of $A^T$ are linearly independent, $U_k$ can get all the $L$ subfiles of all the files from the available $L$ coded subfiles. In short, every user can decode all the $N$ files without any more transmissions. That is, at $M=\frac{N}{L}$, the transmission rate required is zero. Since there is no transmission, the question of secure delivery is meaningless.
	
\begin{figure}[t]
		\begin{center}
			\captionsetup{justification = centering}
			\includegraphics[width = \columnwidth]{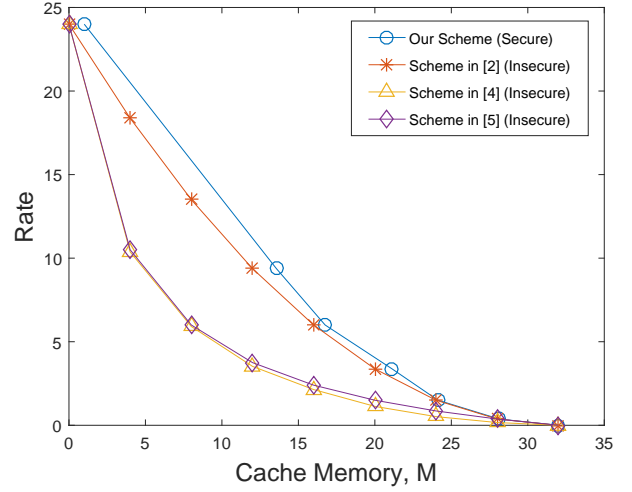}
			\caption{Performance Comparison: $(K=24,L=3,N=96)$ Multi-access Coded Caching Scheme. The schemes in \cite{ReK,ReK2,CLWZC} do not satisfy the security condition \eqref{Security}.}
			\label{plot}
		\end{center}
	\end{figure}

	The in-between memory-rate pairs are securely achievable by memory sharing. We plot the rate-memory trade-off of the proposed scheme for a $(K=24,L=3,N=96)$ multi-access coded caching scheme in Fig. \ref{plot}. Also, the proposed scheme is compared with the multi-access coded caching schemes proposed in \cite{ReK,ReK2,CLWZC}. These schemes do not satisfy the security condition. 
	
\noindent \textit{Proof of Secure Delivery}

We have, \hspace{0.2cm} $I(X^{\mathbf{d}};\mathbf{W}) = H(X^{\mathbf{d}})-H(X^{\mathbf{d}}|\mathbf{W}).$

But,\hspace{0.3cm}	$H(X^{\mathbf{d}})\leq K(1-\frac{iL}{K})^2 \text{ }F \text{ bits}.$
	
   \noindent  Case 1:  At M=1,
	\begin{align*}	
     	H(X^{\mathbf{d}}|\mathbf{W}) &= H(\mathbb{K}_k: k\in [K]|\mathbf{W})\\
     	                             &= H(\mathbb{K}_k: k\in [K])= KF \text{ bits.}
     \end{align*}
        Case 2 ($gcd(K,i) = 1$): At $M=\frac{iN}{K}+\frac{K}{L}(1-\frac{iL}{K})^2$, 
    \begin{align*}	
    	H(X^{\mathbf{d}}|\mathbf{W}) &= H(\mathbb{K}_\alpha: \alpha\in [(K-iL)^2]|\mathbf{W})\\
    	&= H(\mathbb{K}_\alpha: \alpha\in [(K-iL)^2])\\
    	&= K\left(1-\frac{iL}{K}\right)^2 F \text{ bits.}
    \end{align*}
        Case 3 ($gcd(K,i) = g \neq 1$): At $M=\frac{iN}{K}+\frac{2K}{g(L+1)}(1-\frac{iL}{K})^2$,
    \begin{align*}	
	H(X^{\mathbf{d}}|\mathbf{W}) &= H(\mathbb{K}_\alpha: \alpha\in [g(\tilde{K}-\tilde{i}L)^2]|\mathbf{W})\\
	&= H(\mathbb{K}_\alpha: \alpha\in [g(\tilde{K}-\tilde{i}L)^2])\\
	&= K\left(1-\frac{iL}{K}\right)^2 F \text{ bits.}
    \end{align*}
In all the above cases, $H(X^{\mathbf{d}})\leq H(X^{\mathbf{d}}|\mathbf{W})$. Therefore, $I(X^{\mathbf{d}};\mathbf{W}) = 0.$ 

\section{conclusion}
In this work, we have analyzed the multi-access coded caching problem in the presence of an external wiretapper. We have proposed a multi-access coded caching scheme with secure delivery which uses key encryption for each multicast transmission. Also, we have introduced a distributed placement strategy for the keys. The proposed secure multi-access coded caching scheme is optimal within a constant multiplicative gap of 6 when $L\geq \frac{K}{2}$ and $N\geq 2K$. The fraction of extra memory needed for key-placement is almost negligible, especially when number of files with the sever is large. 

\section*{Acknowledgment}
\label{Ack}
This work was supported partly by the Science and Engineering Research Board (SERB) of Department of Science and Technology (DST), Government of India, through J.C. Bose National Fellowship to B. Sundar Rajan.

\appendices
\section{Proof of Theorem \ref{OptimalityGap}}
\label{appA}
Consider a $(K,L,N)$  multi-access coded caching scheme with $L\geq \frac{K}{2}$. $R^*(M)$ is the information-theoretic optimal rate at cache memory $M$, restricted to uncoded placement of file contents. If we restrict to uncoded data placement, the memory-rate pairs $(1,K)$, $(\frac{N}{K}+\frac{(K-L)^2}{KL}, K(1-\frac{L}{K})^2)$ and $(\frac{2N}{K},0)$ are securely achievable. In this case, for $1\leq M\leq \frac{2N}{K}$,
	\begin{equation*}
		\frac{R_s(M)}{R^*(M)} \leq 
		\begin{cases}
			6 & \text{if $2K\leq N < 3K$. }\\
			4  & \text{if $N\geq 3K$.}
		\end{cases}   
	\end{equation*}
To prove the optimality gap, we use the fact that the achievable scheme in \cite{ReK} (Insecure scheme) achieving $R\left(\frac{iN}{K}\right) = K(1-\frac{iL}{K})^2$ is optimal within a factor 2 for $L\geq \frac{K}{2}$. That is, 
\begin{equation*}
	\frac{R(M)}{R^*(M)}\leq 2.
\end{equation*} 
To prove the optimality gap, we need to show that,
\begin{equation}
	\label{bound}
	\frac{R_s(M)}{R(M)} \leq 
	\begin{cases}
		3 & \text{if $2K\leq N < 3K$. }\\
		2  & \text{if $N\geq 3K$.}
	\end{cases}   
\end{equation}
We will show that \eqref{bound} holds at $M=1$, $M=\frac{N}{K}$ and for \\$M\geq \frac{N}{K}+\frac{K}{L}\left(1-\frac{L}{K}\right)^2$. Since, \eqref{bound} holds at the corner points, we can argue that the inequality holds also at the in-between memory points.
\begin{enumerate}
	\item At $M=1$\\
	By memory sharing between the $(M,R)$ pairs $(0,K)$ and $(\frac{N}{K},K\left(1-\frac{L}{K}\right)^2)$, we obtain,
	\begin{equation*}
		R(1) = K-\frac{2LK-L^2}{N}.
	\end{equation*}
Also, $
	R_s(M) = K.$
Therefore,
\begin{align*}
	\frac{R_s(1)}{R(1)} &= \frac{K}{K-\frac{2LK-L^2}{N}}. 
\end{align*}
But, $\frac{2LK-L^2}{N}\leq \frac{K^2}{N}$. So,
\begin{align*}
	\frac{R_s(1)}{R(1)} &\leq  \frac{K}{K-\frac{K^2}{N}}
	                    = \frac{1}{1-\frac{K}{N}}\\
	                    &\leq 2 \hspace{0.5cm} \text{  for } N\geq 2K. 
\end{align*}
	\item At $M = \frac{N}{K}$\\
	By memory sharing between the $(M,R)$ pairs $(1,K)$ and $(\frac{N}{K}+\frac{K}{L}\left(1-\frac{L}{K}\right)^2,K\left(1-\frac{L}{K}\right)^2)$, we obtain,
	\begin{equation*}
		R_s\left(\frac{N}{K}\right) = K-\frac{K(1-\beta)(\frac{N}{K}-1)}{(\frac{N}{K}-1)+\beta \frac{K}{L}}.
	\end{equation*}
	where $\beta := \left(1-\frac{L}{K}\right)^2$. Therefore,
	\begin{align*}
		\frac{R_s\left(\frac{N}{K}\right) }{R\left(\frac{N}{K}\right) } &= \frac{K\left(1-\frac{(1-\beta)(\frac{N}{K}-1)}{(\frac{N}{K}-1)+\beta \frac{K}{L}}\right)}{K\beta} \\
		&= \frac{1-\frac{(1-\beta)(\frac{N}{K}-1)}{(\frac{N}{K}-1)+\beta \frac{K}{L}}}{\beta}
		\leq \frac{1}{\beta}- \frac{\frac{1}{\beta}-1}{1+\frac{2\beta}{t}},
	\end{align*}
     where $t = \frac{N}{K}-1$. We have,
	\begin{align*}
		\frac{R_s\left(\frac{N}{K}\right) }{R\left(\frac{N}{K}\right) } &
		\leq \frac{1+\frac{2}{t}}{1+\frac{2\beta}{t}}\\
		&\leq 1+\frac{2}{t} = 1+\frac{2}{\frac{N}{K}-1}.
	\end{align*}
So, we get,
\begin{equation*}
	\frac{R_s\left(\frac{N}{K}\right) }{R\left(\frac{N}{K}\right) } \leq 
	\begin{cases}
		3 & \text{if $2K\leq N < 3K$. }\\
		2  & \text{if $N\geq 3K$.}
	\end{cases}   
\end{equation*}
\item For $\frac{N}{K}+\frac{K}{L}\left(1-\frac{L}{K}\right)^2\leq M \leq 2K$\\
By memory sharing between $(\frac{N}{K},K\left(1-\frac{L}{K}\right)^2)$ and $(\frac{2N}{K},0)$, we obtain,
\begin{equation*}
	R(M) = K\left(2-\frac{M}{N/K}\right)\left(1-\frac{L}{K}\right)^2.
\end{equation*}
By memory sharing between the memory-rate pairs\\ $(\frac{N}{K}+\frac{K}{L}\left(1-\frac{L}{K}\right)^2,K\left(1-\frac{L}{K}\right)^2)$ and $(\frac{2N}{K},0)$, we have,

\begin{equation*}
	R_s(M) = \frac{K\left(2-\frac{M}{N/K}\right)\left(1-\frac{L}{K}\right)^2}{1-\frac{K/L}{N/K}\left(1-\frac{L}{K}\right)^2}.
\end{equation*}
For, $\frac{N}{K}+\frac{K}{L}\left(1-\frac{L}{K}\right)^2\leq M \leq 2K$,
\begin{align*}
	\frac{R_s(M)}{R(M)} &=  \frac{1}{1-\frac{K/L}{N/K}\left(1-\frac{L}{K}\right)^2}
	                     \leq \frac{1}{1-\frac{1}{2N/K}}.
\end{align*}
For the last inequality, we used the fact that for $L\geq \frac{K}{2}$, $\frac{K}{L}\left(1-\frac{L}{K}\right)^2\leq \frac{1}{2}$. Therefore,
\begin{equation*}
	\frac{R_s(M)}{R(M)} 
	\leq \frac{1}{1-\frac{1}{2N/K}}\leq 2 \hspace{0.3cm} \text{for} N\geq K.
\end{equation*} 
\end{enumerate}
So, in conclusion
\begin{equation*}
	\frac{R_s(M)}{R(M)} \leq 
	\begin{cases}
		3 & \text{if $2K\leq N < 3K$. }\\
		2  & \text{if $N\geq 3K$.}
	\end{cases}   
\end{equation*}
And, for $1\leq M\leq \frac{2N}{K}$,
\begin{equation*}
	\frac{R_s(M)}{R^*(M)} \leq 
	\begin{cases}
		6 & \text{if $2K\leq N < 3K$. }\\
		4  & \text{if $N\geq 3K$.}
	\end{cases}   
\end{equation*}
This completes the proof of Theorem \ref{OptimalityGap}.\hfill $\blacksquare$


\begin{thebibliography}{9}
	\bibitem{MaN} 
M. A. Maddah-Ali and U. Niesen, “Fundamental limits of caching,” \textit{IEEE Trans. Inf. Theory}, vol. 60, no. 5, pp. 2856–2867, May 2014.

\bibitem{ReK}
K. S. Reddy and N. Karamchandani, "Rate-Memory Trade-off for Multi-Access Coded Caching With Uncoded Placement," in \textit{IEEE Transactions on Communications},  vol. 68, no. 6, pp. 3261-3274, June 2020.	

\bibitem{SPE}
B. Serbetci, E. Parrinello and P. Elia, "Multi-access coded caching: gains beyond cache-redundancy," in \textit{IEEE Information Theory Workshop (ITW)}, Visby, Sweden, 2019, pp. 1-5.		

\bibitem{ReK2}
Kota Srinivas Reddy, Nikhil Karamchandani, "Structured Index Coding Problem and Multi-access Coded Caching," arXiv:2012.04705 [cs.IT], Dec 2020.

\bibitem{CLWZC}
Minquan Cheng, Dequan Liang, Kai Wan, Mingming Zhang, Giuseppe Caire, "A Novel Transformation Approach of Shared-link Coded Caching Schemes for Multiaccess Networks," arXiv:2012.04483 [cs.IT], Dec 2020.

\bibitem{SaR}
Shanuja Sasi, B. Sundar Rajan, "Multi-access Coded Caching Scheme with Linear Sub-packetization using PDAs," arXiv:2102.06616 [cs.IT], Feb 2021.


\bibitem{SaR2}
Shanuja Sasi, B. Sundar Rajan, "An Improved Multi-access Coded Caching with Uncoded Placement,"  arXiv:2009.05377v3 [cs.IT], Feb 2021.

\bibitem{MaR}
Anjana A Mahesh, B. Sundar Rajan, "A Coded Caching Scheme with Linear Sub-packetization and its Application to Multi-Access Coded Caching,"  	arXiv:2009.10923 [cs.IT], Sep 2020.

\bibitem{STC}
A. Sengupta, R. Tandon and T. C. Clancy, "Fundamental Limits of Caching With Secure Delivery," in \textit{IEEE Transactions on Information Forensics and Security}, vol. 10, no. 2, pp. 355-370, Feb. 2015.

\bibitem{RPKP}
V. Ravindrakumar, P. Panda, N. Karamchandani and V. M. Prabhakaran, "Private Coded Caching," in \textit{IEEE Transactions on Information Forensics and Security}, vol. 13, no. 3, pp. 685-694, March 2018.


\bibitem{WaG}
K. Wan and G. Caire, "On Coded Caching With Private Demands," in \textit{IEEE Transactions on Information Theory}, vol. 67, no. 1, pp. 358-372, Jan 2021.

\bibitem{Kam}
Sneha Kamath, “Demand Private Coded Caching,” arXiv:1909.03324[cs.IT], Sep 2019.

\bibitem{AST}
V R Aravind, Pradeep Kiran Sarvepalli, Andrew Thangaraj, “Coded Caching with Demand Privacy: Constructions for Lower Subpacketization and Generalizations,” arXiv:2007.07475 [cs.IT], Jul 2020.

\bibitem{GRKDK}
Chinmay Gurjarpadhye, Jithin Ravi, Sneha Kamath, Bikash Kumar Dey, Nikhil Karamchandani, "Fundamental Limits of Demand-Private Coded Caching," arXiv:2101.07127 [cs.IT], Jan 2021.		

\bibitem{NaR}
K.K. Krishnan Namboodiri, B. Sundar Rajan, "Optimal Demand Private Coded Caching for Users with Small Buffers," arXiv:2101.08745v2 [cs.IT] , Feb 2021.	

\bibitem{YaT}
Qifa Yan, Daniela Tuninetti, "Fundamental Limits of Caching for Demand Privacy against Colluding Users," arXiv:2008.03642v1 [cs.IT], Aug 2020.		



\bibitem{YBJK}
Z. Bar-Yossef, Y. Birk, T. S. Jayram and T. Kol, "Index Coding With Side Information," in \textit{IEEE Transactions on Information Theory}, vol. 57, no. 3, pp. 1479-1494, March 2011.

\bibitem{VaR}
M. B. Vaddi and B. S. Rajan, "Optimal Scalar Linear Index Codes for One-Sided Neighboring Side-Information Problems," in \textit{IEEE Globecom Workshops (GC Wkshps)}, Washington, DC, USA, 2016, pp. 1-6.	


\bibitem{Sha}
 C. E. Shannon, “Communication theory of secrecy systems,” \textit{Bell Syst. Tech. J.}, vol. 28, no. 4, pp. 656–715, Sep. 1949.
\end{thebibliography}
\end{document}